\documentstyle[12pt]{article}

\newcommand{\bra}[1]{\langle{#1}|}
\newcommand{\ket}[1]{|{#1}\rangle}

\newcommand{\eq}[1]{Eq.~(\ref{#1})}

\def\beq{\begin{equation}}
\def\eeq{\end{equation}}
\def\beqa{\begin{eqnarray}}
\def\eeqa{\end{eqnarray}}
\newcommand{\sect}[1]{\setcounter{equation}{0}\section{#1}}

\newcommand{\EQ}{\begin{equation}}
\newcommand{\EN}{\end{equation}}
\newcommand{\bea}{\begin{eqnarray}}
\newcommand{\ena}{\end{eqnarray}}

\renewcommand{\a}{\alpha}

\newcommand{\NP}[1]{Nucl.\ Phys.\ {\bf #1}}
\newcommand{\PL}[1]{Phys.\ Lett.\ {\bf #1}}

\newcommand{\PR}[1]{Phys.\ Rev.\ {\bf #1}}
\newcommand{\PRL}[1]{Phys.\ Rev.\ Lett.\ {\bf #1}}

\renewcommand{\thefootnote}{\fnsymbol{footnote}}
\def\one{{\hbox{ 1\kern-.8mm l}}}

\def\NS{{\rm NS}}
\def\R{{\rm R}}
\def\ii{{\rm i}}

\newlength{\bredde}
\def\slash#1{\settowidth{\bredde}{$#1$}\ifmmode\,\raisebox{.15ex}{/}
\hspace*{-\bredde} #1\else$\,\raisebox{.15ex}{/}\hspace*{-\bredde} #1$\fi}

\begin{document}
\begin{titlepage}
\rightline{DFTT 20/98}
\rightline{NORDITA 98/40-HE}
\rightline{KUL-TF-98/25}
\rightline{\hfill May 1998}
\vskip 1.2cm
\centerline{\Large \bf The Lorentz force between D0 and D6 branes} 
\centerline{\Large \bf  in string and M(atrix) theory
\footnote{Work partially supported by the European Commission
TMR programme ERBFMRX-CT96-0045 in which R.R. is associated to
Torino University, and by MURST.}}
\vskip 1.2cm
\centerline{\bf M. Bill\'o$^a$\footnote{e-mail:
billo@tfdec1.fys.kuleuven.ac.be}, P. Di Vecchia$^b$, M. Frau$^{c,f}$,}
\vskip .2cm
\centerline{\bf 
A. Lerda$^{d,c,f}$, R. Russo$^{e,f}$ and S. 
Sciuto$^{c,f}$} \vskip .5cm
\centerline{\sl $^a$ Instituut voor theoretische fysica,}
\centerline{\sl Katholieke Universiteit Leuven, B-3001 Leuven, Belgium}
\vskip .2cm
\centerline{\sl $^b$ NORDITA, Blegdamsvej 17, DK-2100 Copenhagen \O, 
Denmark} \vskip .2cm
\centerline{\sl $^c$ Dipartimento di Fisica Teorica, Universit\`a di
Torino}
\vskip .2cm
\centerline{\sl $^d$ Dipartimento di Scienze e Tecnologie Avanzate}
\centerline{\sl Universit\`a di Torino, sede di Alessandria}
\vskip .2cm
\centerline{\sl $^e$ Dipartimento di Fisica, Politecnico di Torino}
\vskip .2cm
\centerline{\sl $^f$ I.N.F.N., Sezione di Torino, Via P. Giuria 1, 
I-10125 Torino, Italy}
\vskip 1.5cm
\begin{abstract}
We use different techniques to analyze the system formed by a D0 brane
and a D6 brane (with background gauge fields) in relative motion.
In particular, using the closed string formalism
of boosted boundary states, we show the presence of a
term linear in the velocity, corresponding to the Lorentz force
experienced by the D0 brane moving in the magnetic background produced 
by the D6 brane. This term, that was missed in previous analyses of
this system, comes entirely from the R-R odd spin structure and 
is also  reproduced by a M(atrix) theory calculation.
\end{abstract}
\end{titlepage}
\newpage
\renewcommand{\thefootnote}{\arabic{footnote}}
\setcounter{footnote}{0}
\setcounter{page}{1}
\sect{Introduction}
\label{intro}

The interaction between two D-branes \cite{POLCH}
can be computed in many different
ways: from the string point of view, the short-distance physics
is entirely described by the open strings stretched between
the branes \cite{DKPS}, while, from the M(atrix) theory point of view,
it is reproduced by super Yang-Mills calculations in 0+1
dimensions \cite{BANKS} \footnote{For a review of the role of gauge theories 
in the D-brane physics see Ref.~\cite{TAYLOR} and 
references therein.}. On the other 
hand, the long-distance behavior
of D-branes is conveniently captured by the closed string boundary state
formalism \footnote{A detailed discussion of the boundary state formalism can be
found in Ref.~\cite{BILLO} and
references therein.}, where the D-branes are described by BRST
invariant states 
and the interaction is computed by connecting them with closed string
propagators \cite{CANGEMI,FRAU,cpb,BILLO}.

In this paper we use different techniques to analyze the system
formed by a static D6 brane and a D0 brane moving along a direction
transverse to the D6 brane \cite{BACHAS}.
In all cases, we find that the interaction contains a linear term in
the velocity that was missed in previous analyses from
both the string theory and the M(atrix) theory side
\cite{LIFS,PIERRE,LARSEN,KRAUS}.
However, this term has a very simple classical interpretation: in fact,
since a D0 and a D6 brane are Hodge dual to each other, they are
the analogue in ten dimensions of pointlike electric and
magnetic charges in four dimensions; thus a velocity dependent
interaction is expected, since it represents the Lorentz force
experienced by a particle (the D0 brane) moving in a magnetic background
(produced by the D6 brane).
The presence of this term, in a similar physical
situation, has been recently pointed out in Ref.~\cite{BIS}, where
the authors studied the off-diagonal interaction between a D3 brane and
a D3 brane-antibrane system using string techniques. 
In particular, they have related the magnetic
force to the presence of the odd spin structure of the R-R sector
that usually gives no contribution.

This remark suggests that there should be a relation between a system of
a D0 and a D6 brane in relative motion, and the one formed by a D0 and
a D8 brane at rest.
In fact also in the latter case the odd spin structure gives a non
vanishing contribution, as it happens, in general, for the interaction
between two parallel D-branes when the difference of their 
dimensionalities  $\nu$ is $8$. Since these systems satisfy the
BPS no-force condition \cite{POLCH}, a non vanishing R-R contribution
is needed to compensate the gravitational
interactions of the NS-NS sector. On the other hand, 
a R-R interaction between
two D-branes of different dimensionality is difficult to understand
because the two D-branes seem to couple to different
R-R potentials. However, as explained in Ref.~\cite{BILLO}, the R-R
interaction in the $\nu=8$ brane systems arises entirely from the  
odd spin structure where the presence of a non-trivial
regulator on the matter and superghost zero-modes leads to duality
relations between different R-R potentials.
As we shall see, a similar mechanism can explain the
non-vanishing R-R interaction between a D6 brane and a
moving D0 brane. On the other hand, from the M(atrix) theory point of
view, the term corresponding to the odd spin structure in string
computation is due to the presence of an unpaired eigenvector of the
fermionic lagrangian \cite{LI,HO} which is a feature common to both the 
D0-D6 and D0-D8 brane systems.

In order to make the tight connection between the two systems
immediately evident, let us start by describing them at low energies in
the source-probe framework \cite{PS}. 
The key ingredient in this approach is that
different R-R form potentials may describe the same physical field and 
should
be identified. This can be seen by studying in detail the massless R-R
sector in the asymmetric picture as was recently done in
Ref.~\cite{BILLO}, where it was shown that the Hilbert space in this 
sector contains two dimensional subspaces with degenerate metric:
removing the null-states, one is led to identify pairs of different
vectors. Normally this identification reduces to the usual Hodge duality
for the R-R potentials, but the string analysis reveals
that, for off-shell states, there are also non-trivial relations.
In particular, the R-R potentials of a one-form
coupled to a D0 brane and of a nine-form coupled to a D8 brane
with non-vanishing momentum in the 9th direction are identified:
\beq
A_0 = - A_{012 \dots 8}~.
\label{ide453}
\eeq
Using this result, the presence of a
R-R interaction in the D0-D8 system is interpreted as due to the fact 
that the R-R charges of the two branes are identified by
a duality relation, and thus produce the same R-R potential.
According to this analysis, the R-R
interaction of the D0-D8 system has then a simple microscopic
description as the usual Coulomb-like force between D-branes.
As a check on the validity of this interpretation, we now show
that \eq{ide453} implies that the R-R interaction in the D0-D8 brane
system leads to the static potential
\beq
V_{08}(r) = \frac{1}{4 \pi \alpha '}\,r~,
\label{interr}
\eeq
where
$r$ is the distance between the two branes,
whose presence explains the by-now well known phenomenon
of the creation of a fundamental string
when the D0 brane passes through the D8 brane \cite{GABE}.
Let us consider the low-energy effective action of a
D$0$ brane with tension $\tau_0$ and charge $\mu_0$ in presence of a R-R 
background, which in the static gauge reads
\beq
S_{eff} = - \tau_{0} \int d\tau \sqrt{1 - ({\dot{x}}^i)^2}-  \mu_0 \int
d\tau \left(A_0 + {\dot{x}}^i A_i\right)~.
\label{boinf}
\eeq
Then, if we use the identification (\ref{ide453}), we can view the D8 
brane as a source for the potential experienced by the D0 brane, so that
\beq
A_0 = -A_{012 \dots 8} = \frac{\mu_8}{2}\, r~,
\label{pote08}
\eeq
where $\mu_p = \sqrt{2\pi} (2 \pi \alpha ')^{3-p}$ is the coupling of a
$(p+1)$-form R-R field to a D$p$ brane. Inserting the potential 
(\ref{pote08}) into the Wess-Zumino part of the Born-Infeld action 
(\ref{boinf}) and using the identity $\mu_0 \mu_8 = \frac{1}{2 \pi 
\alpha'}$, we see that the term proportional to $A_0$ immediately
reproduces \eq{interr}.
This simple argument provides a nice consistency check 
of our identification of the potentials produced by a zero and an eight
brane to explain the appearance of the term in \eq{interr}.

In order to apply a similar procedure to the D0-D6 brane system,
one has to find whether there exists some relation between the R-R 
potentials generated by a D6 brane and those coupled to a moving D0
brane. For the sake of simplicity, we suppose that the world
volume of the D6 brane lies in the directions (0)123456, and that the D0
brane velocity $v$ is in the 9th direction with an impact 
parameter ${\vec b}$ in the (78) plane. 
In this configuration, the standard Hodge duality relations read 
$\partial_7A_{012\dots 6}=\partial_8A_9$ and
$\partial_8A_{012\dots 6}=-\partial_7A_9$, where
$A_{012\dots 6}$ is the seven-form emitted by the D6 brane and
$A_9$ is the component of the one-form coupled to the moving D0 brane.
Then, using the explicit expression of $A_{012\dots 6}$, we can write
\begin{equation} A_9({\vec b}, v\tau) = -
\int^{\vec b}(dx^7\partial_8 - dx^8\partial_7)\left(
\frac{\mu_6}{4\pi}\frac{1}{r}\right)~~,
\label{a9}
\end{equation}
where we have defined $r^2={\vec x}\cdot {\vec x}+v^2\tau^2$.
Inserting this relation inside the D0 brane effective action and 
recalling that $\mu_0\mu_6=2\pi$, 
the last term of \eq{boinf} becomes
\beq
-\int^{\vec b} (dx^7\partial_8 - dx^8\partial_7)
\left(\int_{-\infty}^\infty d\tau\,V_{06}(r)
\right)
\label{pote654}
\eeq
where we have defined the velocity-dependent long-range 
``magnetic'' potential 
\beq
V_{06}(r) = -\frac{v}{2r}~,
\label{potlor}
\eeq
which, through \eq{pote654}, describes a Lorentz-like interaction.

In Section 2, using  boundary states and following the procedure
discussed in detail in Ref.~\cite{BILLO}, we compute the interaction
between a D6 brane (with a background magnetic field) at rest and a 
moving D0  brane. In particular we focus on the R-R sector 
where we expect to find from the odd spin structure
a non vanishing contribution of the type (\ref{potlor}).
In Section 3, we consider the same system from the M(atrix) theory point
of view and show that the same term coming from the odd
spin structure in string computation is obtained from a
contribution of the fermionic determinants.
At first sight, this agreement between the two different approaches may
appear surprising: in fact, the M(atrix) theory results usually agree
with those obtained by supergravity calculations only in the infinite
momentum frame and thus it seems that there is no reason to
explain why, for the odd spin structure contribution, the long distance
result coming from the boundary state should always coincide with the
short distance M(atrix) calculation. However, in Section 2, performing
the calculation of the R-R sector, we will find that in the odd-spin
structure, all the stringy massive states give no contribution; this
property is not peculiar of the D0-D6 brane system, but is a general
feature common to all the interactions between two arbitrary boundary
states. The fact that the whole result comes from the zero-modes of the
various fields ensures that the behavior is the same at all scales.
In fact, in order to deduce the short distance behavior from the results
written in the closed string channel, one should usually make the modular
transformation $t \rightarrow 1/t$; however, when the massive modes give
no contribution, this transformation simply maps the contribution of the
massless string states into that of the massless open string and thus for
the odd-spin structure the ``duality'' between loops and tree
diagrams holds also at the field theory level.

\sect{String theory calculation}
\label{boundarystate}
In the boundary state formalism, a D$p$ brane is described by a BRST 
invariant state $\ket{B}$, which inserts a boundary on the 
string world-sheet and represents the source for the closed strings
emitted by the brane.
As discussed in Ref.~\cite{BILLO}, the boundary state both in
the NS-NS and in the R-R sector of the fermionic string
can be written as the product of a matter part and a ghost part
\begin{equation}
\label{bs001}
\ket{B, \eta } = \frac{T_p}{2}\ket{B_{\rm mat}, \eta } \ket{B_{\rm g}, 
\eta}~~, 
\end{equation}
where $\eta=\pm1$, 
$T_p= \sqrt{\pi}\left(2\pi\sqrt{\alpha'}\right)^{3-p}$
is the brane tension, and
\begin{equation}
\ket{B_{\rm mat}, \eta} = \ket{B_X} 
\ket{B_{\psi}, \eta}~~~,~~~
\ket{B_{\rm g}, \eta} = \ket{B_{\rm gh}} \ket{B_{\rm sgh}, \eta}~~.
\label{bs000}
\end{equation}
The explicit expressions of the various components of the boundary state
are given in Ref.~\cite{BILLO}. Here, we simply recall the general 
structure of the matter sector \footnote{The ghost contribution is the 
same as in Ref. \cite{BILLO}, and we do not rewrite it here again.}, 
namely 
\begin{equation} \label{bs100}
\ket{B_X} = \exp\biggl[-\sum_{n=1}^\infty \frac{1}{n}\,
\a_{-n}\cdot S \cdot
\tilde \a_{-n}\biggr]\,
\ket{B_X}^{(0)}~~,
\end{equation}
for the bosonic part, and
\begin{equation}
\label{bs101}
\ket{B_\psi,\eta}_{\rm NS} =  \exp\biggl[\ii\eta\sum_{m=1/2}^\infty
\psi_{-m}\cdot S \cdot \tilde \psi_{-m}\biggr]
\,\ket{0}~~,
\end{equation}
\begin{equation}
\label{bs102}
\ket{B_\psi,\eta}_\R = \exp\biggl[\ii\eta\sum_{m=1}^\infty
\psi_{-m}\cdot S \cdot \tilde \psi_{-m}\biggr]
\,\ket{B_\psi,\eta}_\R^{(0)}~~,
\end{equation}
respectively for the NS-NS and R-R sectors of the fermionic part.
In these expressions, the matrix $S_{\mu\nu}$
encodes the boundary conditions characterizing the D$p$ brane which  
depend on its velocity and the presence of gauge fields
on its world-volume. The superscript $^{(0)}$ 
in Eqs. (\ref{bs100}) and (\ref{bs102}) denotes 
the zero-mode contributions which also depend on the boundary conditions. 
Let us now give some details for the configuration
we want to study.  The matrix $S$ for a D$0$ brane 
moving with a velocity $v$ along the 9-th direction, 
is given by the 
following block-diagonal form
\begin{equation}  
\label{sd0}
S= {\rm diag}(-{\cal V}_{09}, -\one_{12}, -\one_{34}, -\one_{56},
-\one_{78})
\end{equation}
where $\one_{ij}$ is the identity in the $i$-th and $j$-th directions,
and ${\cal V}_{09}$ is a $2\times 2$ matrix, acting in 
the $0$-th and $9$-th directions, given by
\beq
{\cal V}_{09} =  \left(\begin{array}{cc}
          \frac{1+v^2}{1-v^2}  & \frac{2v}{1 - v^2} \\
         \frac{2v}{1-v^2}  & \frac{1+ v^2}{1-v^2}
\end{array} \right)
= \left(\begin{array}{cc}
          \cosh(2\pi\nu)  & \sinh(2\pi\nu) \\
         \sinh(2\pi\nu)  & \cosh(2\pi\nu)
\end{array} \right)~,
\label{emab}
\eeq
where
\begin{equation}
v = \tanh (\pi \nu)~~.
\label{vnu}
\end{equation}
This result can be derived by applying the boost operator
$\exp ({\rm i}\pi\nu J^{09})$  ($J^{\rho\sigma}$ being the
generators of the Lorentz transformations) to the boundary state
of a D$0$ brane at rest, as explicitly done in Ref. \cite{CANGEMI}.
The action of the boost operator on the zero-mode part of 
the boundary state yields for the bosonic sector
\begin{equation}
\label{zerox}
\ket{B_X}^{(0)} = \sqrt{1 - v^2} \delta^{(8)}( \hat q - y) 
\delta \left( {\hat{q}}^0 v +  {\hat{q}}^9  \right)
\ket{k=0}~~,
\end{equation}
where $\delta^{(8)}$ denotes 
the $\delta$-function that fixes the position of the 
D0 brane in all transverse directions except the $9$-th along which
the motion takes place.
In the fermionic R-R sector, instead, one
finds
\beq
\label{bsr012}
\ket{B_\psi,\eta}_\R^{(0)} =
{\cal M}_{AB}^{(\eta)}\,\ket{A} \ket{\tilde B}~~, 
\eeq
where
\begin{equation}
\label{bs1453}
{\cal M}^{(\eta)} = C\Gamma^0
\left( \frac{1- v \Gamma^{09}}{\sqrt{1-v^2}} \right)
\left(\frac{1+\ii\eta\Gamma_{11}}{1+\ii\eta}\right)~~,
\end{equation}
with $C$ being the charge conjugation matrix. 

Let us now turn to the D$6$ brane. To be able to compare the string
results with the M(atrix) theory calculation performed in the next 
section, we
switch on a constant magnetic field on the world volume of a static
D$6$ brane. In this way, we actually describe a 
$(6+4+2+0)$-bound state.
The matrix $S$ corresponding to this configuration is
given by the 
following block-diagonal form
\begin{equation}  
\label{sd6}
S= {\rm diag}(-\one_{09}, {\cal F}_{12}, 
{\cal F}_{34}, {\cal F}_{56}, -\one_{78})~,
\end{equation}
where the $2\times 2$ matrix ${\cal F}_{ij}$,
acting in the $i$-th and $j$-th directions, is 
\footnote{For simplicity we have taken the same value of $f$ in all 
three blocks. The generalization to different values is 
straightforward.} 
\beq {\cal F}_{ij} =  \left(\begin{array}{cc}
          \frac{1-f^2}{1+f^2}  & \frac{2f}{1 +f^2} \\
         -\frac{2f}{1+f^2}  & \frac{1-f^2}{1+f^2}
\end{array} \right)
= \left(\begin{array}{cc}
          \cos(2\pi\epsilon)  & \sin(2\pi\epsilon) \\
         -\sin(2\pi\epsilon)  & \cos(2\pi\epsilon)
\end{array} \right)~~,
\label{emabf}
\eeq
where
\begin{equation}
f = \tan(\pi\epsilon)~~.
\label{fepsi}
\end{equation}
This result can be obtained by generalizing the construction
described in Ref.\cite{cpb}, where it was shown how to
create a bound state of D-branes by performing a rotation
followed by a T-duality transformation. Correspondingly, 
the zero-mode piece of the bosonic boundary state is found to be
\begin{equation}
\label{zerox6}
\ket{B_X}^{(0)} =  (\sqrt{1+f^2})^3
\delta^{(3)}( \hat q - y) 
\ket{k=0}~~,
\end{equation}
where the delta function fixes the position of the D$6$ brane 
in its three transverse directions ({\it i.e.} 7,8,9), whereas
for the R-R sector it is like in \eq{bsr012} with the
matrix ${\cal M}^{(\eta)}$ given by
\begin{equation}
{\cal M}^{(\eta)} =
C\Gamma^{01...56}\left[\left(\frac{1- f 
\Gamma^{12}}{\sqrt{1+f^2}}\right)
\left(\frac{1- f \Gamma^{34}}{\sqrt{1+f^2}}\right)
\left(\frac{1- f \Gamma^{56}}{\sqrt{1+f^2}}\right)\right]
\left(\frac{1+\ii \eta \Gamma_{11}}{1+\ii\eta}\right)~.
\label{rr6}
\end{equation}
This matrix clearly exhibits
the structure of a bound state of D6, D4, D2 and D0 branes,
corresponding to the terms with zero, two, four and six $\Gamma$
matrices in the square brackets, and is formally similar to
the boosted matrix in \eq{bs1453}.

We are now in the position of computing the scattering amplitude
\beq
{\cal{A}} = \bra{B_1} D \ket{B_2}~,
\label{matel}
\eeq
where $D$ is the closed string propagator, and $\ket{B_1}$
and $\ket{B_2}$ are the GSO projected boundary states describing
the moving D0 brane and the D6 brane bound state respectively.
Let us begin by discussing the NS-NS contribution to ${\cal A}$. 
Proceeding as explained in detail in Refs.~\cite{BILLO,CANGEMI}
we get 
\bea
{\cal{A}}_{\NS-\NS} 
&=&\frac{\sqrt{1-v^2}}{32\sqrt{2\pi\alpha'}}
\int_{- \infty}^{\infty} d \tau
\int_{0}^{\infty} \frac{dt}{t^{3/2}}~ e^t 
\exp \left[ - \frac{( \Delta y)^2 + v^2 \tau^{2}}{2 \alpha ' t}\right]
\nonumber \\
&\times&
\left\{ \prod_{n=1}^{\infty} 
\frac{\det ( 1 + q^{2n-1} S_2 (S_1)^T)}{\det ( 1 - q^{2n} 
S_2 (S_1)^T)}
\frac{(1 - q^{2n})^2}{(1 + q^{2n-1})^2} \right.
\label{NSNS}\\
&&~-\left.
\prod_{n=1}^{\infty} 
\frac{\det ( 1 - q^{2n-1} S_2 (S_1)^T)}{\det ( 1 - 
q^{2n} S_2 (S_1)^T)}
\frac{(1 - q^{2n})^2}{(1 - q^{2n-1})^2} \right\}~,
\nonumber 
\ena
where $q={\rm e}^{-t}$, and $S_1$ and $S_2$ are given 
respectively by Eqs. (\ref{sd0}) and (\ref{sd6}). The last two lines
of \eq{NSNS} represent the contribution of the non-zero modes
of the NS-NS closed string states exchanged between the two D-branes.
The two terms in braces correspond to the two NS-NS spin
structures and the relative minus sign assures the cancellation of
the tachyon. We remark that to compute this part,
it is not necessary to specify the detailed form of the 
matrices $S$, thus the non-zero mode contribution
always takes this form for any brane
configuration. Upon inserting the explicit expressions for $S_1$ and
$S_2$, and rescaling the modular parameter $t \to \pi t$, one
can rewrite \eq{NSNS} in terms of Jacobi $\theta$ functions
and get
\bea
{\cal{A}}_{NS-NS} &=& \frac{\,v}{2\pi\sqrt{2\alpha'}}
\int_{- \infty}^{\infty} d \tau
\int_{0}^{\infty} \frac{dt}{t^{3/2}} ~
\exp \left[- \frac{r^2}{2 \pi\alpha ' t}\right]
\nonumber \\
&\times&
\left\{ \ii\,\frac{\theta_{3} ( \ii\nu | \ii t)}
{\theta_{1} ( \ii\nu | \ii t)} 
\left( \frac{\theta_{4} ( \epsilon | \ii t)}
{\theta_{2} ( \epsilon | \ii t) }\right)^{3} - 
\ii\,\frac{\theta_{4} ( \ii\nu | \ii t)}
{\theta_{1} ( \ii\nu | \ii t)} 
\left(\frac{\theta_{3} ( \epsilon | \ii t)}
{\theta_{2} ( \epsilon | \ii t) }\right)^{3}
\right\}~,
\label{NSNS1}
\ena
where $r^2=(\Delta y)^2+v^2\tau^2$, and the arguments
of the $\theta$ functions are given by Eqs. (\ref{vnu}) 
and (\ref{fepsi}). 

Let us now turn to the R-R sector. In this case it is easier to
perform the calculation before the GSO projection. After some 
straightforward algebra we find
\bea
{}_\R\bra{B^1 , \eta_1 } D \ket{B^2 , \eta_2}_\R
&=& \frac{1}{16\sqrt{2\pi\alpha'}}
\int_{- \infty}^{\infty}  d \tau
\int_{0}^{\infty} \frac{dt}{t^{3/2}}~
\exp \left[ - \frac{r^2}{2 \alpha ' t}\right]
\label{RR1} \\
&\times&
\sqrt{1-v^2}\left(\sqrt{1+f^2}\right)^3 \,
{}^{(0)}_\R \langle{B^1 , \eta_1} \ket{B^2 , \eta_2}^{(0)}_\R
\nonumber \\
&\times&
\left\{\delta_{\eta_1 \eta_2 , 1} +  \delta_{\eta_1 \eta_2 ,- 1} 
\prod_{n=1}^{\infty} 
\frac{\det ( 1 + q^{2n} S_2 (S_1)^T)}{\det ( 1 - q^{2n} S_2 (S_1)^T)}
\frac{(1 - q^{2n})^2}{(1 + q^{2n})^2} \right\}~.
\nonumber \ena
In this expression 
${}^{(0)}_\R \langle{B^1 , \eta_1} 
\ket{B^2 , \eta_2}^{(0)}_\R$
denotes the zero-mode contribution, while the
terms in braces are due to the non-zero modes of the R-R states 
exchanged between the two D-branes. Notice that in
the piece proportional to $\delta_{\eta_1\eta_2,1}$ there is an exact 
cancellation between the bosonic and fermionic non-zero modes  
while they survive in the piece proportional to
$\delta_{\eta_1\eta_2,-1}$. This cancellation is an algebraic fact due 
to world-sheet supersymmetry, and always occurs for any D-brane
configuration, {\it  i.e.} independently of the explicit expressions of 
the matrices $S_1$ and $S_2$. Since, as one can see, the term 
proportional to $\delta_{\eta_1\eta_2,1}$ corresponds to the odd R-R 
spin structure, while the term proportional to 
$\delta_{\eta_1\eta_2,-1}$
corresponds to the even R-R spin structure, we expect a very different
behavior of the respective interactions. We will return to this crucial
point later.

To evaluate the zero mode contribution in \eq{RR1} one has to follow the
procedure described in detail in Ref. \cite{BILLO}, which in particular
requires not to separate the fermionic and superghost zero-modes and to
use a regulator inside the inner product 
\footnote{The fermionic regulator that must be used in 
this case is
$x^{\Gamma^{09}}x^{\Gamma^{12}}x^{\Gamma^{34}}
x^{\Gamma^{56}}x^{\Gamma^{78}}$. This expression, which slightly
differs from the one used in Ref.~\cite{BILLO}, namely
$x^{\Gamma^{09}}x^{-\ii\Gamma^{12}}x^{-\ii\Gamma^{34}}
x^{-\ii\Gamma^{56}}x^{-\ii\Gamma^{78}}$, yields a well-defined
and unitary inner-product also in those cases, like the present one, in
which the ghosts and superghosts effectively remove two spatial
directions ({\em e.g.} the 7th and 8th).}. 
Following this
procedure, it is 
easy to find that \beq
{}^{(0)}_{\R} \langle {B^{1}_{\psi}, \eta_1}  
\ket{B^{2}_{\psi} , \eta_2}^{(0)}_{\R}=-16\, \frac{1}{\sqrt{1-v^2}}
\left(\frac{1}{\sqrt{1+f^2}}\right)^3\left(f^3 \delta_{\eta_1\eta_2,-1}
-v \,\delta_{\eta_1 \eta_2 , 1}\right)~.
\label{zero987}
\eeq
Inserting this result  in \eq{RR1}, using the explicit expressions for 
the matrices $S_1$ and $S_2$, performing the GSO projection and 
introducing the Jacobi $\theta$ functions, we obtain
\bea
{\cal{A}}_{\R-\R} 
&=&\frac{v}{2\pi\sqrt{2\alpha'}}
\int_{- \infty}^{\infty} d \tau
\int_{0}^{\infty} \frac{dt}{t^{3/2}} ~
\exp \left[- \frac{r^2}{2 \pi\alpha ' t}\right]
\nonumber \\
&\times&
\left\{ -\ii\,\frac{\theta_{2} ( \ii\nu | \ii t)}
{\theta_{1} ( \ii\nu | \ii t)} 
\left( \frac{\theta_{1} ( \epsilon | \ii t)}
{\theta_{2} ( \epsilon | \ii t) }\right)^{3} +1\right\}~.
\label{RR12}
\ena
Performing the modular transformation $t\to 1/t$ one can check that
our results (\ref{NSNS1}) and (\ref{RR12}) agree with the phase 
shift calculated in Ref. \cite{LIFS,PIERRE}
using the open string formalism, except for the odd spin structure
contribution which was missed in the previous
calculations. It is interesting to remark that the same expression
(\ref{RR12}) can be obtained using the light-cone methods described in
Ref.~\cite{MORALES}, thus providing a strong consistency check on our
results. On the other hand, our analysis clearly shows, the 
configuration of a D0 brane
scattering off a D6 brane has the same R-R zero-mode
content as the D0-D8 brane system, where it is known 
that the odd spin structure
is non vanishing \cite{GABE,BILLO}. On the other hand, 
the presence of a contribution from the odd spin structure
and its physical relevance have been recently discussed in Ref. 
\cite{BIS} for a system similar to the one we are considering.

{}From the total amplitude ${\cal{A}}_{\NS-\NS}+{\cal{A}}_{\R-\R}$, we 
can now define the long range potential between the two D-branes
due to the exchange of the massless closed string states. To do so, we 
treat separately the even and odd spin structures, since 
they correspond to different types of 
interactions. Indeed, the even spin
structures describe the usual interplay between 
the gravitational attraction
from the NS-NS sector and the Coulomb-like electric 
repulsion from the R-R sector, while the
odd spin structure accounts for the Lorentz-like 
electro-magnetic interaction.
Expanding Eqs. (\ref{NSNS1}) and (\ref{RR12}) 
for $t\to\infty$ in order to
select the contribution of the massless closed string 
states, we can introduce an even and odd  
long range potential given respectively by
\begin{equation} 
V_{{\rm even}}(r)
= -\frac{v}{8}\left(\frac{\cosh{2\pi\nu}-3\cos{2\pi\epsilon}-4
\cosh{\pi\nu}~\sin^3{\pi\epsilon}}{\sinh{\pi
\nu}~ \cos^3{\pi\epsilon}} \right) \frac{1}{r}
\label{poteven}
\end{equation}
and
\begin{equation}
V_{{\rm odd}}(r) = -\frac{v}{2r}~.
\label{potodd}
\end{equation}
The potential $V_{\rm even}$ agrees with the one computed in
Ref. \cite{PIERRE},
whereas the potential $V_{\rm odd}$ 
coincides with the one in \eq{potlor} 
obtained with the probe-source approach. Besides a different 
physical interpretation, the two potentials
(\ref{poteven}) and (\ref{potodd}) have 
also a very different behavior. Indeed, $V_{\rm odd}$
has the same $r$ dependence at all scales, because it 
arises from the
odd R-R spin structure in which
the non-zero modes exactly cancel.
This peculiar feature guarantees that the potential at large
distances, where only the lightest closed string modes contribute,
is {\it exactly} the same as the potential at small distances 
where only the lightest open string modes contribute. 
For this reason, one should expect that the Lorentz-like
potential (\ref{potodd}) be reproduced also 
from a M(atrix) theory calculation. 
On the other hand, the potential $V_{\rm even}$
arises from the even spin structures in which the non-zero
modes do not cancel. Thus, the long distance potential
(\ref{poteven}) is different from the short distance one obtained
by truncating the open string to its massless level.
However, there is a regime in which the non-zero modes
can be neglected also for the even spin structures so that 
an agreement between the long and short-distance 
behaviors can be obtained, 
namely for small velocities and high fields. 
To discuss this regime, it
is convenient to set $\epsilon = 1/2 - c/\pi$ 
and expand the phase shift of the even spin structures for 
$\nu,c\to 0$. Then, one can explicitly check from
Eqs. (\ref{NSNS1}) and (\ref{RR12}) that
only the massless closed string
states contribute to the leading
order in $\nu$ and $c$, whereas the massive modes give
contributions of higher order. Thus, only the leading order term in the
expansion of \eq{poteven}, namely
\begin{equation}
\label{potevenapp}
V_{\rm even} (r) \simeq -\frac{v^4+6v^2c^2-3c^4}{16\,c^3}\,\frac{1}{r}~,
\end{equation}
where $c\simeq 1/f$, represents a potential valid
at all distances. Indeed, the same expression arises from
a M(atrix) theory calculation \cite{LIFS,PIERRE,LARSEN,KRAUS} in
the infinite momentum frame. In the following section, we show that
the agreement between the closed string calculation and the M(atrix)
theory actually occurs, as it should, also for the Lorentz-like potential
(\ref{potodd}) without making any approximation or 
choosing the infinite momentum frame.

\sect{M(atrix) theory calculation}

As usual, the starting point for M(atrix) theory calculations is the 10D 
lagrangian of super Yang--Mills reduced to 1+0 dimensions;
however, in
this section, we focus only on the fermionic part of this 
lagrangian because our purpose is to find the term corresponding
to the odd spin structure which can only come from the fermionic
sector. On the other hand, the bosonic sector of this M(atrix) 
theory has been analyzed and discussed in Ref. \cite{LIFS,PIERRE}.
Thus we just consider the following lagrangian \cite{TAYLOR}
\beq
\label{l1+0}
L_f = {1\over 2}{\rm Tr} \left(\bar \psi\, \ii \Gamma^0 
\partial_t \psi - \bar \psi \Gamma^i
[X_i, \psi] \right)~,
\eeq
where $X^i$ are the bosonic degrees of freedom, while
$\psi$ are 10-dimensional Majorana-Weyl spinors.
In the case of a two D-brane system, these fields are $U(2)$
matrices with well defined classical values. We can describe
a system of a D0 and a D6 brane with relative velocity $v$ in 
the 9th direction and impact parameter $b$ in the 8th direction,
by using the following background matrices for the bosonic
fields
\beq\label{backg}
X_{2i-1} =  \left(\begin{array}{cc}
Q_i & 0 \\ 0 &0
\end{array}\right)~,
\hspace{.5cm}
X_{2i} = \left(\begin{array}{cc}
P_i & 0 \\ 0 & 0
\end{array}\right)~,
\hspace{.5cm}
\eeq
$$
X_7 = 0~, \hspace{.5cm}
X_8 = \left( \begin{array}{cc}
b & 0 \\ 0 &0
\end{array} \right)~,
\hspace{.5cm}
X_9 = \left( \begin{array}{cc}
0 & 0 \\ 0 & vt
\end{array} \right)~,
$$
where $i=1,2,3$, and
\beq\label{comm}
[Q_i,P_j] = \ii c_i \delta_{ij}~,
\eeq
and trivial classical values for the fermionic variables.
Note that the above choice for the background fields 
does not describe a pure D-brane, but a D6 brane with a constant
magnetic field switched on, that is a $(6+4+2+0)$ bound state, just
like the one discussed in the previous section. 
Thus, besides the magnetic contribution, that is linear in the velocity,
the M(atrix) theory result will also contain an electric term due 
to the interaction between the moving D0 brane and the D0 branes
contained in the bound state. However, it is easy to separate the two
contributions since the magnetic force is independent of $c_i$, that are
related to the gauge fields switched on the D6 brane.

Now we choose an explicit representation for
$\Gamma^\mu$, by taking 
$\Gamma^0=\ii\sigma_1\otimes\gamma^8$,
$\Gamma^i=\sigma_1\otimes
\gamma^i$ $(i=1,...,6)$, $\Gamma^7=\sigma_2\otimes\one$,
$\Gamma^8=\sigma_1\otimes\gamma^9$, $\Gamma^9=\sigma_1\otimes
\gamma^7$, where $\gamma^i$ are the eight
$16\times 16$ gamma matrices of $SO(8)$ and $\gamma^9=\gamma^1\cdots
\gamma^8$. The reason for this unusual choice is simply a matter of 
later convenience, since with this representation the D0-D6 problem 
becomes formally similar to the D0-D8 problem studied in Ref.~\cite{LI}.
Indeed, using  \eq{backg}, 
the lagrangian (\ref{l1+0}) can be rewritten as follows
\bea\label{lf1+0}
L_f &=& - \frac{\ii}{2}\left(\chi^{T}\,\partial_t \chi +
\xi^{T}\,\partial_t \xi\right) - \ii\, 
\bar\theta\,\left[\gamma^8\partial_t + Q_1\gamma^1
+ P_1\gamma^2 + Q_2\gamma^3 + P_2\gamma^4\right.
 \\ \nonumber  & &\left.+
Q_3\gamma^5 + P_3\gamma^6 + b\,\gamma^9
-v t\,\gamma^7\right]\theta = - {\ii\over 
2}\left(\chi^{T}\,\partial_t\chi
+ \xi^{T}\, \partial_t\xi\right)-\ii\,\bar\theta\,D_\theta 
\,\theta ~,
\ena
where the 16 component spinors $\chi$ and $\xi$ (real) and $\theta$ 
(complex) describe the fermionic fluctuations of the Majorana-Weyl 
spinor $\psi$ arranged according to
\beq\label{ffluc}
\delta\psi =  \left(\begin{array}{cc}
\chi & \theta \\ \theta^\dagger & \xi
\end{array}\right)~.
\eeq
In \eq{lf1+0} $\bar \theta=\theta^\dagger \gamma^8$ as 
follows from the specific representation of $\Gamma^0$ we 
have chosen. Just like in the 
open string formalism, the first contribution to the
interaction between two D-branes is given by the one-loop vacuum energy 
$E$. In our case only the off-diagonal fluctuations of $\psi$ give a 
non-vanishing result, since the fields $\chi$ and $\xi$ are massless, and 
thus the whole calculation reduces to the evaluation of the functional 
determinant of $D_\theta$. It is easy to see from \eq{lf1+0} that 
the lagrangian for the field $\theta$ is very similar to the one 
encountered in the computation of the static interaction between 
a D0 and a D8 brane. In fact, in this latter case, one 
chooses, for the bosonic degrees of freedom, a background with
four pairs of operators satisfying the commutation relations 
(\ref{comm}); however, in the D0-D6 brane scattering, the velocity 
formally plays the same role of a gauge field $c_i$, since the background
values for $X^9$ and $X^{0}$ are proportional to non-commuting operators
of the same type of those appearing in the other directions.
In order to make the analogy more evident, we perform a Wick rotation 
and introduce the euclidean ``velocity'' v$\,=-\ii v$, so that the
$\theta$-part of the lagrangian becomes
\beq
L_\theta  = -\ii\,\bar\theta\left[\ii\gamma^8\partial_{\rm t} + 
Q_1\gamma^1\! + P_1\gamma^2\! + Q_2\gamma^3\! + P_2\gamma^4\! +
Q_3\gamma^5\! + P_3\gamma^6\! + b\gamma^9\!
-{\rm v}\,{\rm t}\gamma^7\right]\!\theta~.
\eeq
Now we can introduce a fourth pair of operators $Q_4=-{\rm v\,t}$ and 
$P_4=\ii \partial_{\rm t}$ such that
\beq\label{Q5P5}
[Q_4\,,P_4] =  \ii\,{\rm v}~,
\eeq
and thus the operator $D_\theta$ can be written as
\bea\label{llag}
D_{\theta} & = & \left[b\,\gamma_9 +~\slash{m}\right]~,
\\ \nonumber
\slash{m} &=& \left[ Q_1\gamma^1 + P_1\gamma^2 + 
Q_2\gamma^3 + P_2\gamma^4+ Q_3\gamma^5 + P_3\gamma^6 + 
Q_4\gamma^7 + P_4\gamma^8\right]~.
\ena
This lagrangian was studied in detail in Refs. \cite{LI,HO} and here 
we will simply recall its main feature: from each eigenvector of 
$D_\theta$ ($D_\theta\ket{V}_r = \lambda_r\ket{V}_r$) that is 
not chiral ($\gamma^9 \ket{V} \not= \pm\ket{V}$) another eigenvector with
opposite eigenvalue can be constructed; on the contrary $D_\theta$ has 
only one chiral eigenvector. In fact, it is easy to check that 
$\ket{V}_{-r}
= (b-\lambda_r\gamma^9) \ket{V}_r$ is also an eigenvector of $D_\theta$ 
with eigenvalue $(-\lambda_r)$. However, this construction fails if
$\gamma^9 \ket{V}_c=\pm\ket{V}_c$: in this case, $~\slash{m}$ must vanish
when it acts on $\ket{V}_c$ (because it always changes the chirality of a
vector) and $\lambda_c$ is simply equal to $\pm b$, thus, the combination
$\ket{V}_{-c}$ vanishes. If, for the sake of simplicity, we suppose that 
the background values are positive ($c_i,{\rm v}>0$), this unpaired
eigenvector $\ket{V}_c$ is chiral $\gamma^9\ket{V}_c = \ket{V}_c$. The
presence of the zero-mode of $~\slash{m}$ makes the evaluation of the 
determinant of $D_\theta$ a bit delicate. In fact, as it was pointed out 
in Ref.~\cite{HO}, when $b$ is zero the chiral eigenvector is a 
zero-mode for the full lagrangian and thus must be projected out in the 
definition of the determinant. Assuming that this projection has to
be done also when $b$ is adiabatically switched on, we compute
the effective energy as follows
\beq\label{ve}
E \equiv {\rm Tr'}\left[ \ln(-\ii\,D_\theta)\right] ={\rm Tr} 
\left[ \left( {1-\gamma^9 \over 2}\right) \ln(b^2+\slash{m}^2)\right]~,
\eeq
where the identity $b^2+\slash{m}^2 = (\ii D_\theta)(-\ii D_\theta)$ and
the pairing of the non chiral eigenvectors of $D_\theta$ have been used.
Using the explicit expression of $\slash{m}$ and the 
commutation relations of the operators $Q_i$ and $P_i$, we obtain 
\beq
E= \sum_{N_1,N_2,N_3,N_4=\pm1}
{\rm tr} \left[\left(\frac{1-N_1N_2N_3N_4}{2}\right)
\log\left(b^2+\sum_{i=1}^4 H_i - \sum_{i=1}^3 c_i N_i
-{\rm v} N_4\right)\right]
\label{step}
\eeq
where $N_i=-\ii \gamma^{2i-1}\gamma^{2i}$, and $H_i=P_i^2+Q_i^2$.
The sum over $N_i$'s represents the trace over the 16 dimensional
spinor space, and  
the symbol ${\rm tr}$ in \eq{step} means the trace over the 
eigenstates of the harmonic oscillator hamiltonians $H_i$, whose 
spectrum is $2c_i(n_i+1/2)$ for $i=1,2,3$ and $2{\rm v}(n_4+1/2)$
for $i=4$. Then, using the ``Schwinger'' parametrization for the
logarithm and summing over $n_i$, we get
\begin{equation}
 \label{n13}
 E = - \int_0^\infty {ds \over s} {\rm e}^{-sb^2} \,
 {1\over 2}\left[ \left(\prod_{i=1}^3  {\cosh c_i s\over \sinh c_i 
 s}\right) {\cosh {\rm v}s\over \sinh {\rm v}s} - 1\right]~.
\end{equation}
By undoing the Wick rotation, the first term in the square
bracket together with the contribution of the bosonic determinants
computed for example in Ref.~\cite{LIFS}, correctly reproduces the
even part of the interaction potential. On the contrary, 
the last term in the square bracket of \eq{n13}, which is 
independent of the gauge fields $c_i$ of the D6 brane, arises because
of the insertion of the chiral projector and accounts for the 
Lorentz-like magnetic interaction. By the usual 
trick
\beq\label{trick}
1 = v \sqrt{s\over \pi}\int_{-\infty}^\infty {\rm e}^{-s 
(v\tau)^2}\,d\tau ~, 
\eeq
we can reintroduce the velocity dependence and verify that this term 
reproduces the same contribution that in string theory calculation
came from the R-R odd spin structure. In fact, 
\begin{equation}
 \label{matr}
E_{odd} ={v\over 2\sqrt{\pi}} \int_{-\infty}^\infty \!\! 
d\tau
\int_0^\infty {ds \over s^{1/2}} {\rm e}^{-s(b^2+(v\tau)^2)} =
\frac{v}{2} \int_{-\infty}^{\infty} \frac{1}{r}\,d \tau ~, 
\end{equation}
leading to the same potential of Eqs. (\ref{potlor}) and (\ref{potodd}). 

It is interesting to see what happens if one includes also the 
contribution of the chiral zero-mode of $\slash{m}$, thus computing the 
full one-loop effective action $\Gamma$ of the matrix model.
To expose the effect more clearly, it is convenient to consider a
general background in which both $X_7$ and $X_8$ are constant and different
from zero. In this case, the impact parameter becomes a vector ${\vec 
b}$ with components $b_7$ and $b_8$, and the operator $D_\theta$ in 
\eq{llag} gets replaced by $D_\theta=-\ii b_7 \one + b_8\gamma_9 
+\slash{m}$. Using the properties we have 
mentioned above, it is not difficult to see that 
\bea
\Gamma &\equiv& {\rm Tr}\left[ \ln(-\ii\,D_\theta)\right] =
{\rm Tr'}\left[ \ln(-\ii\,D_\theta)\right] 
+ \ln\left(b_7 +\ii b_8
\right) 
\nonumber \\ &=& 
{1 \over 2}\left( {\rm  Tr}[\ln({\vec b}^{\,2}+\slash{m}^2)] - 
\ln {\vec b}^{\,2}\right)
+\ln\left(b_7 +\ii b_8\right) 
\nonumber \\ &=&
- \int_0^\infty {ds \over s} {\rm e}^{-s {\vec b}^{\,2}} \,
 {1\over 2}\left(\prod_{i=1}^3  {\cosh c_i s\over \sinh c_i 
 s}\right) {\cosh {\rm v}s\over \sinh {\rm v}s} +
\ii\,{\rm Im}\ln\Big(b_7+\ii b_8\Big)~.
\label{Gamma}
\ena
The two terms of this expression represent the even and the 
odd parts of the effective action which account respectively for  
Coulomb-like and Lorentz-like interactions of the D0-D6 brane system.
Notice that the last term in \eq{Gamma} exactly coincides with the
expression in \eq{pote654} obtained with the probe-source formalism,
and correctly describes the phase-shift produced by a Lorentz-like
interaction. 

In conclusion we have shown both in string and in M(atrix) theory that
the interaction between between pairs of Hodge dual D-branes in relative
motions contains a term due to the Lorentz force.

 \vskip 1cm \noindent
{\large{\bf {Acknowledgements}}}
\vskip 0.5cm
\noindent
We thank M. Bertolini, I. Pesando and C. A. Scrucca for very useful 
discussions, and R. Iengo for useful remarks.


\begin{thebibliography}{99}

\bibitem{POLCH} J. Polchinski, Phys. Rev. Lett {\bf 75} (1995) 4724,
{\tt hep-th/9510017}; J. Polchinski, S. Chaudhuri and C.V. Johnson, {\it
Notes on D-branes}, {\tt hep-th/ 9602052};
J. Polchinski, {\it TASI lectures on D-branes}, 
{\tt hep-th/9611050}.

\bibitem{DKPS}  M. R. Douglas, D. Kabat, P. Pouliot, S. H. Shenker, 
\NP{B485} (1997) 85, {\tt hep-th/9608024}.

\bibitem{BANKS} T. Banks, W. Fischler, S. H. Shenker, L. Susskind,
\PR{D55} (1997) 5112, {\tt hep-th/9610043}.

\bibitem{TAYLOR} A. Bilal, {\it M(atrix) Theory : a Pedagogical 
Introduction}, {\tt hep-th/9710136}; D. Bigatti and L. Susskind, 
{\it Review of Matrix Theory}, {\tt hep-th/9712072}; W. Taylor, {\em
Lectures on D-branes, gauge theory and M(atrices)}, {\tt hep-th/
9801182}.

\bibitem{BILLO} M. Bill{\'{o}}, P. Di Vecchia, M. Frau, A. Lerda, I. Pesando,
R. Russo and S. Sciuto, {\em Microscopic string analysis of the D0-D8
brane system and dual R-R states}, {\tt hep-th/9802088}.

\bibitem{CANGEMI} 
M. Bill{\'{o}}, D. Cangemi and P. Di Vecchia, \PL{400B} (1997)
63, {\tt hep-th/9701190}.

\bibitem{FRAU} M. Frau, I. Pesando, S. Sciuto, A. Lerda and R. Russo,
\PL{400B} (1997) 52, {\tt hep-th/9702037}.

\bibitem{cpb}P. Di Vecchia, M. Frau, I.
Pesando, S. Sciuto, A. Lerda, R. Russo, \NP{B507} (1997) 259, 
{\tt hep-th/9707068}.

\bibitem{BACHAS} C. Bachas, \PL{374B} (1996) 37, {\tt hep-th/9511043}.
G. Lifschytz, \PL{388B} (1996) 720, {\tt hep-th/9604156} 

\bibitem{LIFS} G. Lifschytz, {\em Four brane and six brane interactions 
in M(atrix) theory}, {\tt hep-th/9612223}.

\bibitem{PIERRE} J. M. Pierre, Phys.Rev. {\bf D56} (1997) 6710,
{\tt hep-th/9707102}.

\bibitem{LARSEN} V. Balasubramanian, F. Larsen, \NP{B506} (1997) 61
{\tt hep-th/9703039}.

\bibitem{KRAUS} E. Keski-Vakkuri, P. Kraus, \NP{B510} (1997) 199,
{\tt hep-th/9706196}.

\bibitem{BIS} M. Bertolini, R. Iengo and C. A. Scrucca, {\em
Electric and magnetic interaction of dyonic D-branes and odd spin
structure}, {\tt hep-th/9801110}.

\bibitem{LI} P. M. Ho, M. Li and Y. S. Wu, {\em $p-p'$ strings in 
M(atrix) theory}, {\tt hep-th/ 9706073}.

\bibitem{HO} P. M. Ho and Y. S. Wu, {\em Brane creation in M(atrix) 
theory}, {\tt hep-th/9708137}.

\bibitem{PS} I. Chepelev, A. A. Tseytlin, {\em Long-distance interactions
of branes: correspondence between supergravity and super Yang-Mills
descriptions}, {\tt hep-th/ 9709087}.

\bibitem{MORALES}J. F. Morales, C. A. Scrucca and M. Serone, \PL{417B} 
(1998) 233, {\tt hep-th/9709063}; 
{\em Scale independent spin-effects in D-brane dynamics}, {\tt 
hep-th/9801183}.
 
\bibitem{GABE} C. P. Bachas, M. R. Douglas and  M. B. Green, 
JHEP {\bf 07} (1997) 002, {\tt hep-th/ 9705074};
U. Danielsson, G. Ferretti and I. R. Klebanov, \PRL{79} (1997)
1984, {\tt hep-th/ 9705084};
O. Bergman, M. Gaberdiel and G. Lifschytz,
\NP{B509} (1998) 194, {\tt hep-th/9705130}.
\end{thebibliography}
\end{document}